\documentclass[]{aastex701}
\usepackage{graphicx}

\begin{document}

\title{Ultra-Deep imaging of the starless galaxy candidate Cloud-9}

\author[orcid=0000-0001-8647-2874]{Ignacio Trujillo}
\affiliation{Instituto de Astrof\'{i}sica de Canarias, V\'{i}a L\'{a}ctea S/N, E-38205 La Laguna, Spain}
\affiliation{Departamento de Astrof\'{i}sica, Universidad de La Laguna, E-38206 La Laguna, Spain}
\email[show]{trujillo@iac.es}  

\author[orcid=0009-0003-6502-7714]{Ignacio Ruiz Cejudo}
\affiliation{Instituto de Astrof\'{i}sica de Canarias, V\'{i}a L\'{a}ctea S/N, E-38205 La Laguna, Spain}
\affiliation{Departamento de Astrof\'{i}sica, Universidad de La Laguna, E-38206 La Laguna, Spain}
\email{ignaciomrrd.es@gmail.com}

\author[orcid=0009-0001-7407-2491]{Sergio Guerra Arencibia} 
\affiliation{Instituto de Astrof\'{i}sica de Canarias, V\'{i}a L\'{a}ctea S/N, E-38205 La Laguna, Spain}
\affiliation{Departamento de Astrof\'{i}sica, Universidad de La Laguna, E-38206 La Laguna, Spain}
\email{sguerra@iac.es}

\author[orcid=0000-0001-7847-0393]{Mireia Montes}
\affiliation{Institute of Space Sciences (ICE, CSIC), Campus UAB, Carrer de Can Magrans, s/n, 08193 Barcelona, Spain}
\email{mmontes@ice.csic.es}


\begin{abstract}

We report the non-detection of stellar light associated with the dark galaxy candidate Cloud-9. Ultra-deep imaging obtained with HiPERCAM at the Gran Telescopio Canarias (GTC) reaches surface-brightness limits of 31.4 and 31.0 mag/arcsec$^2$ in g and r, respectively, approximately ten times deeper than previous deep imaging of the region. No stellar emission is detected within the central 1$\arcmin\times1\arcmin$
 ($\sim$1.3$\times$1.3 kpc) region of Cloud-9. Assuming an old, metal-poor stellar population, these limits imply a stellar surface mass density of no more than $\sim$0.01 M$_\odot$/pc$^2$ and  an upper limit on the total stellar mass of 1.6$\times$10$^4$ M$_\odot$. This independent constraint from integrated light complements previous limits based on resolved-star counts and strengthens the case for Cloud-9 as a candidate starless galaxy.

\end{abstract}


\section{Introduction} 

A direct consequence of the $\Lambda$ Cold Dark Matter scenario is the presence of a large number of low-mass \citep[$\sim$10$^{9.7}$ M$_\odot$ at z=0; ][]{2020MNRAS.498.4887B} dark matter halos with gas but no stars. A balance between gravitational confinement, gas cooling, and heating by the ultraviolet background would prevent the gas in these low-mass dark matter halos from producing stars \citep[][]{2006MNRAS.371..401H,2016MNRAS.456...85S}.  Cloud-9 (a potential satellite of the galaxy M94) is a promising candidate for this type of object  \citep[][]{2023ApJ...952..130Z}. This gas cloud has an observed line width  of W$_{50}$$\lesssim$ 20 km s$^{-1}$ and an H$_{I}$ mass of $\sim$10$^{6}$ M$_\odot$ \citep[][]{2023ApJ...952..130Z,2024ApJ...973...61B}. 

Alternative interpretations have been proposed for starless H$_{I}$ clouds, including tidal debris, high-velocity clouds, and transient gaseous structures. Some may host very diffuse stellar counterparts \citep[][]{2013LNP...861..327D}, motivating ultra-deep imaging of Cloud-9 to place stringent limits on the presence of stars. Previous searches \citep[][]{2023ApJ...952..130Z,2025ApJ...993L..55A} have placed upper limits on its stellar mass, but both have limitations.  \citet[][]{2023ApJ...952..130Z} use the DESI Legacy Survey, whose pipeline is not designed to preserve the low surface brightness (LSB) signal.  \citet[][]{2025ApJ...993L..55A}, using star-counting techniques based on deep Hubble Space Telescope (HST) data, heavily relies on the comparison with the stellar density of the galaxy Leo T   \citep[$\mu_{0,V}$$\sim$27 mag arcsec$^{-2}$ and effective radius around 180 pc; ][]{2007ApJ...656L..13I}. Leo T has a stellar mass of 10$^{5}$ M$_\odot$ but it is relatively compact for its absolute magnitude. Galaxies with similar  luminosities in the Local Group can be four to five times larger (e.g. CVn I, Sex I, etc) and, therefore, two to three magnitudes fainter in surface brightness \citep[][]{2012AJ....144....4M}. For this reason, we  perform ultra-deep imaging of Cloud-9 (with a limiting surface brightness of $\sim$31 mag arcsec$^{-2}$; V-band)  to  independently constrain both its surface brightness and total stellar mass.

\section{GTC ultra deep imaging}

Ultra-deep images of Cloud-9 were obtained using HiPERCAM/GTC \citep{2018SPIE10702E..0LD} in five  filters (u, g, r, i, z) simultaneously. Each of the five CCDs covers 2.7$\times$1.4 arcmin$^2$. To speed up the camera's readout, the detector was binned to a 2$\times$2 pixel mode resulting in a final pixel scale of  
$0.16\,\rm{arcsec\,pixel^{-1}}$. Cloud-9  was observed on  June 11 and 12, 2026 with a seeing  around 0.9\arcsec. We followed a dithering strategy similar to the one described in \citet{2021A&A...654A..40T} and the data reduction follows the steps detailed in \citet{Nube}. We performed photometric calibration of these images using Pan-STARRS1 \citep{panstarrs} for g, r, i and z; and Sloan Digital Sky Survey \citep[SDSS, ][]{sdss} for u.  A constant background was subtracted from each masked frame before coaddition. In addition to masking the foreground and background sources, we masked a circular region with a radius of 215 arcsec centered on Cloud-9 to avoid over-subtracting the light associated with its potential stellar emission. The final exposure time on-source at the center of the image is 2.36h for each band and the limiting surface brightnesses, measured in areas equivalent to 10$\times$10 arcsec$^2$, are 30.1, 31.4, 31.0, 30.4, and 29.6 mag arcsec$^{-2}$ for u, g, r, i, and z respectively (3$\sigma$ above the background).

\begin{figure*}[ht!]
\includegraphics[width=\textwidth]{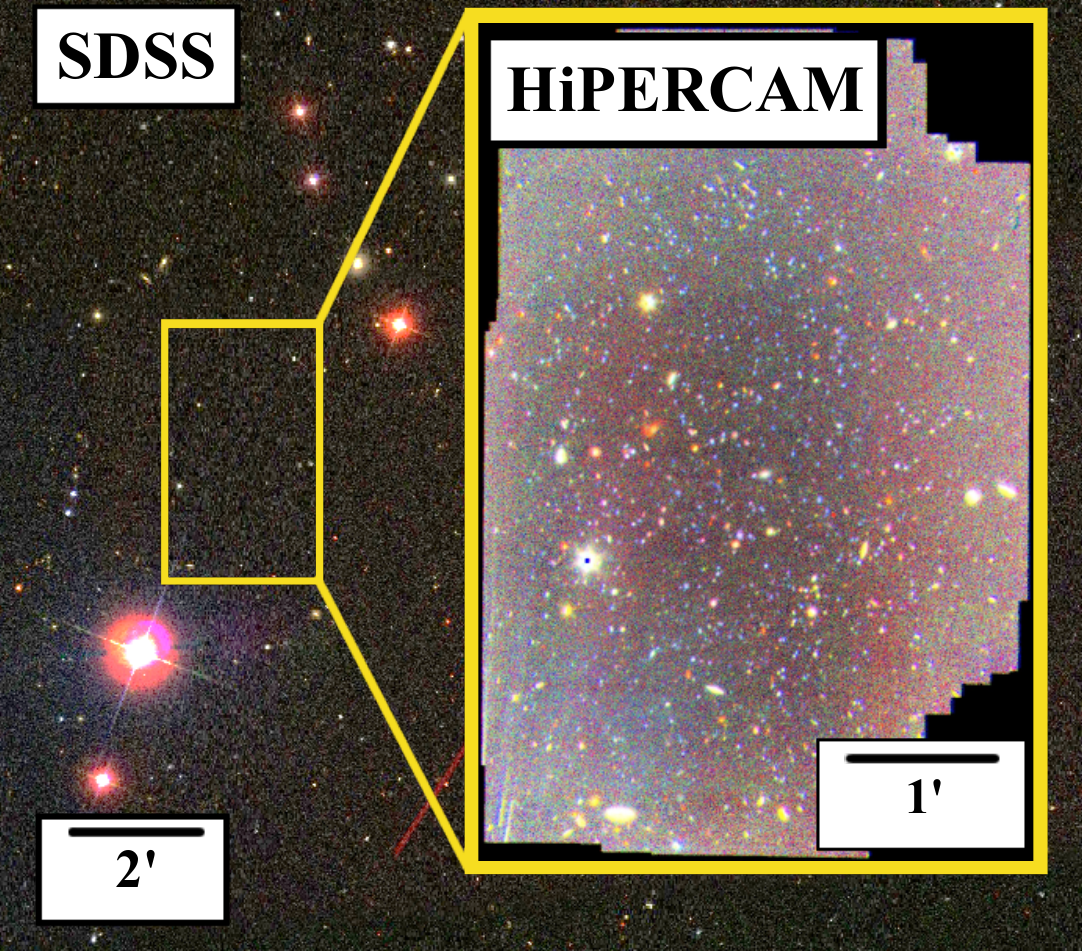}
\caption{Color gri-band composition of the region containing Cloud-9. The SDSS image is shown with a zoomed-in HiPERCAM ultra-deep image centered on Cloud-9. Scattered light from the bright star HD 111859 (V=8.09 mag), located just outside the HiPERCAM field of view at bottom left, affects the lower-left corner of the ultra-deep image. No stellar emission is detected at the location of Cloud-9 down to 31.4 mag/arcsec$^2$ (g-band). Increased noise at the image edges, caused by the large dithering pattern, appears as a whitening of the background.
\label{fig:image}}
\end{figure*}

The lack of any observable features brighter than 31.4 (g-band) and 31.0 (r-band)  mag/arcsec$^2$ in the central 1\arcmin$\times$1\arcmin\ region of the coadded image \citep[equivalent to 1.3 kpc $\times$ 1.3 kpc at the distance of M94, i.e. 4.66 Mpc; ][]{2009AJ....138..332J} imposes  upper limits on the stellar surface density in that region and, therefore, on the total stellar mass enclosed in that area. We assume an old, metal-poor stellar population with an age of 10 Gyr and [Fe/H]=-2 \citep{2025ApJ...993L..55A}. A stellar population with these characteristics has a color of g-r$\sim$0.45 \citep{2010MNRAS.404.1639V} and (M$_{\star}$/L)$_V$=0.86 \citep{2015MNRAS.452.3209R}. Therefore, a limiting surface brightness of 31 mag/arcsec$^2$ (V-band) corresponds to $\Sigma_\star$$\sim$0.01 M$_\odot$/pc$^2$ \citep[][]{2008ApJ...683L.103B} and an upper limit on the total stellar mass of 1.6$\times$10$^4$ M$_\odot$. This independent constraint lies between the two existing upper limits published in the literature: 10$^{5.1}$  M$_\odot$ \citep{2023ApJ...952..130Z} based on deep imaging from the DESI Legacy Survey  and 10$^{3.5}$  M$_\odot$ \citep{2025ApJ...993L..55A} based on deep HST imaging. This new,  upper limit on stellar mass, which is based solely on ultra-deep imaging, supports the idea that Cloud-9 is an excellent candidate for a starless galaxy.

\begin{acknowledgments}
Based on observations obtained with the GTC, installed at the Spanish Observatorio del Roque de los Muchachos of the Instituto de Astrofísica de Canarias on La Palma. IT acknowledges support from the State Research Agency (AEI-MCINN) of the Spanish Ministry of Science and Innovation under grant PID2022-140869NB-I00. This work also acknowledges support from the European Union through the EU Horizon Europe Widening Actions “UNDARK” and “Excellence in Galaxies – Twinning the IAC” (101159929 and 101158446), and MSCA EDUCADO (GA 101119830).
\end{acknowledgments}

\begin{contribution}

IT wrote the Research Note and led the interpretation of the data. IRC and SGA processed and generated the images shown. MM contributed to interpretation of the data.

\end{contribution}




\bibliography{sample701.bib}{}

@INPROCEEDINGS{2018SPIE10702E..0LD,
       author = {{Dhillon}, Vik and {Dixon}, Simon and {Gamble}, Trevor and {Kerry}, Paul and {Littlefair}, Stuart and {Parsons}, Steven and {Marsh}, Thomas and {Bezawada}, Naidu and {Black}, Martin and {Gao}, Xiaofeng and {Henry}, David and {Lunney}, David and {Miller}, Christopher and {Dubbeldam}, Marc and {Morris}, Timothy and {Osborn}, James and {Wilson}, Richard and {Casares}, Jorge and {Mu{\~n}oz-Darias}, Teo and {Pall{\'e}}, Enric and {Rodriguez-Gil}, Pablo and {Shahbaz}, Tariq and {de Ugarte Postigo}, Antonio},
        title = "{First light with HiPERCAM on the GTC}",
    booktitle = {Ground-based and Airborne Instrumentation for Astronomy VII},
         year = 2018,
       editor = {{Evans}, Christopher J. and {Simard}, Luc and {Takami}, Hideki},
       series = {Society of Photo-Optical Instrumentation Engineers (SPIE) Conference Series},
       volume = {10702},
        month = jul,
          eid = {107020L},
        pages = {107020L},
          doi = {10.1117/12.2312041},
archivePrefix = {arXiv},
       eprint = {1807.00557},
 primaryClass = {astro-ph.IM},
       adsurl = {https://ui.adsabs.harvard.edu/abs/2018SPIE10702E..0LD}
}

@ARTICLE{2021A&A...654A..40T,
       author = {{Trujillo}, Ignacio and {D'Onofrio}, Mauro and {Zaritsky}, Dennis and {Madrigal-Aguado}, Alberto and {Chamba}, Nushkia and {Golini}, Giulia and {Akhlaghi}, Mohammad and {Sharbaf}, Zahra and {Infante-Sainz}, Ra{\'u}l and {Rom{\'a}n}, Javier and et al.},
        title = "{Introducing the LBT Imaging of Galactic Halos and Tidal Structures (LIGHTS) survey. A preview of the low surface brightness Universe to be unveiled by LSST}",
      journal = {\aap},
         year = 2021,
        month = oct,
       volume = {654},
          eid = {A40},
        pages = {A40},
          doi = {10.1051/0004-6361/202141603},
archivePrefix = {arXiv},
       eprint = {2109.07478},
 primaryClass = {astro-ph.GA},
       adsurl = {https://ui.adsabs.harvard.edu/abs/2021A&A...654A..40T}
}

@ARTICLE{2020MNRAS.498.4887B,
       author = {{Benitez-Llambay}, Alejandro and {Frenk}, Carlos},
        title = "{The detailed structure and the onset of galaxy formation in low-mass gaseous dark matter haloes}",
      journal = {\mnras},
         year = 2020,
        month = nov,
       volume = {498},
       number = {4},
        pages = {4887-4900},
          doi = {10.1093/mnras/staa2698},
archivePrefix = {arXiv},
       eprint = {2004.06124},
 primaryClass = {astro-ph.GA},
       adsurl = {https://ui.adsabs.harvard.edu/abs/2020MNRAS.498.4887B}
}

@ARTICLE{2006MNRAS.371..401H,
       author = {{Hoeft}, Matthias and {Yepes}, Gustavo and {Gottl{\"o}ber}, Stefan and {Springel}, Volker},
        title = "{Dwarf galaxies in voids: suppressing star formation with photoheating}",
      journal = {\mnras},
         year = 2006,
        month = sep,
       volume = {371},
       number = {1},
        pages = {401-414},
          doi = {10.1111/j.1365-2966.2006.10678.x},
archivePrefix = {arXiv},
       eprint = {astro-ph/0501304},
 primaryClass = {astro-ph},
       adsurl = {https://ui.adsabs.harvard.edu/abs/2006MNRAS.371..401H}
}

@ARTICLE{2016MNRAS.456...85S,
       author = {{Sawala}, Till and {Frenk}, Carlos S. and {Fattahi}, Azadeh and {Navarro}, Julio F. and {Theuns}, Tom and {Bower}, Richard G. and {Crain}, Robert A. and {Furlong}, Michelle and {Jenkins}, Adrian and {Schaller}, Matthieu and {Schaye}, Joop},
        title = "{The chosen few: the low-mass haloes that host faint galaxies}",
      journal = {\mnras},
         year = 2016,
        month = feb,
       volume = {456},
       number = {1},
        pages = {85-97},
          doi = {10.1093/mnras/stv2597},
archivePrefix = {arXiv},
       eprint = {1406.6362},
 primaryClass = {astro-ph.CO},
       adsurl = {https://ui.adsabs.harvard.edu/abs/2016MNRAS.456...85S}
}

@ARTICLE{2023ApJ...952..130Z,
       author = {{Zhou}, Ruilei and {Zhu}, Ming and {Yang}, Yanbin and {Yu}, Haiyang and {Yuan}, Lixia and {Jiang}, Peng and {Xi}, Wenzhe},
        title = "{FAST Reveals New Evidence for M94 as a Merger}",
      journal = {\apj},
         year = 2023,
        month = aug,
       volume = {952},
       number = {2},
          eid = {130},
        pages = {130},
          doi = {10.3847/1538-4357/acdcf5},
archivePrefix = {arXiv},
       eprint = {2306.05080},
 primaryClass = {astro-ph.GA},
       adsurl = {https://ui.adsabs.harvard.edu/abs/2023ApJ...952..130Z}
}

@ARTICLE{2024ApJ...973...61B,
       author = {{Ben{\'\i}tez-Llambay}, Alejandro and {Dutta}, Rajeshwari and {Fumagalli}, Michele and {Navarro}, Julio F.},
        title = "{Examining the Nature of the Starless Dark Matter Halo Candidate Cloud-9 with Very Large Array Observations}",
      journal = {\apj},
         year = 2024,
        month = sep,
       volume = {973},
       number = {1},
          eid = {61},
        pages = {61},
          doi = {10.3847/1538-4357/ad65d9},
       adsurl = {https://ui.adsabs.harvard.edu/abs/2024ApJ...973...61B}
}

@INCOLLECTION{2013LNP...861..327D,
       author = {{Duc}, Pierre-Alain and {Renaud}, Florent},
        title = "{Tides in Colliding Galaxies}",
    booktitle = {Lecture Notes in Physics, Berlin Springer Verlag},
         year = 2013,
       editor = {{Souchay}, Jean and {Mathis}, St{\'e}phane and {Tokieda}, Tadashi},
       volume = {861},
        pages = {327},
          doi = {10.1007/978-3-642-32961-6_9},
       adsurl = {https://ui.adsabs.harvard.edu/abs/2013LNP...861..327D}
}

@ARTICLE{2025ApJ...993L..55A,
       author = {{Anand}, Gagandeep S. and {Ben{\'\i}tez-Llambay}, Alejandro and {Beaton}, Rachael and {Fox}, Andrew J. and {Navarro}, Julio F. and {D'Onghia}, Elena},
        title = "{The First RELHIC? Cloud-9 is a Starless Gas Cloud}",
      journal = {\apjl},
         year = 2025,
        month = nov,
       volume = {993},
       number = {2},
          eid = {L55},
        pages = {L55},
          doi = {10.3847/2041-8213/ae1584},
archivePrefix = {arXiv},
       eprint = {2508.20157},
 primaryClass = {astro-ph.GA},
       adsurl = {https://ui.adsabs.harvard.edu/abs/2025ApJ...993L..55A}
}

@ARTICLE{2012AJ....144....4M,
       author = {{McConnachie}, Alan W.},
        title = "{The Observed Properties of Dwarf Galaxies in and around the Local Group}",
      journal = {\aj},
         year = 2012,
        month = jul,
       volume = {144},
       number = {1},
          eid = {4},
        pages = {4},
          doi = {10.1088/0004-6256/144/1/4},
archivePrefix = {arXiv},
       eprint = {1204.1562},
 primaryClass = {astro-ph.CO},
       adsurl = {https://ui.adsabs.harvard.edu/abs/2012AJ....144....4M}
}

@ARTICLE{Nube,
       author = {{Montes}, Mireia and {Trujillo}, Ignacio and {Karunakaran}, Ananthan and {Infante-Sainz}, Ra{\'u}l and {Spekkens}, Kristine and {Golini}, Giulia and {Beasley}, Michael and {Cebri{\'a}n}, Maria and {Chamba}, Nushkia and {D'Onofrio}, Mauro and {Kelvin}, Lee and {Rom{\'a}n}, Javier},
        title = "{An almost dark galaxy with the mass of the Small Magellanic Cloud}",
      journal = {\aap},
         year = 2024,
        month = jan,
       volume = {681},
          eid = {A15},
        pages = {A15},
          doi = {10.1051/0004-6361/202347667},
archivePrefix = {arXiv},
       eprint = {2310.12231},
 primaryClass = {astro-ph.GA},
       adsurl = {https://ui.adsabs.harvard.edu/abs/2024A&A...681A..15M}
}

@ARTICLE{2008ApJ...683L.103B,
       author = {{Bakos}, Judit and {Trujillo}, Ignacio and {Pohlen}, Michael},
        title = "{Color Profiles of Spiral Galaxies: Clues on Outer-Disk Formation Scenarios}",
      journal = {\apjl},
         year = 2008,
        month = aug,
       volume = {683},
       number = {2},
        pages = {L103},
          doi = {10.1086/591671},
archivePrefix = {arXiv},
       eprint = {0807.2776},
 primaryClass = {astro-ph},
       adsurl = {https://ui.adsabs.harvard.edu/abs/2008ApJ...683L.103B}
}

@ARTICLE{2009AJ....138..332J,
       author = {{Jacobs}, Bradley A. and {Rizzi}, Luca and {Tully}, R. Brent and {Shaya}, Edward J. and {Makarov}, Dmitry I. and {Makarova}, Lidia},
        title = "{The Extragalactic Distance Database: Color-Magnitude Diagrams}",
      journal = {\aj},
         year = 2009,
        month = aug,
       volume = {138},
       number = {2},
        pages = {332-337},
          doi = {10.1088/0004-6256/138/2/332},
archivePrefix = {arXiv},
       eprint = {0902.3675},
 primaryClass = {astro-ph.CO},
       adsurl = {https://ui.adsabs.harvard.edu/abs/2009AJ....138..332J}
}

\bibliographystyle{aasjournal}



\end{document}